\documentclass[preprint,aps,superscriptaddress]{revtex4}

\usepackage{graphics}
\usepackage{dcolumn}
\usepackage{bm}
\usepackage{bbold}
\usepackage{amsmath}
\usepackage{hyperref}
\usepackage{scalerel}
\usepackage{mathtools}
\usepackage{changes}
\usepackage{newtxtext}

\begin{document}


\title{Numerical Computation of Spin-Transfer Torques for Antiferromagnetic Domain Walls}

\author{Hyeon-Jong Park}
\affiliation{KU-KIST Graduate School of Converging Science and Technology, Korea University, Seoul 02841, Korea}

\author{Yunboo Jeong}
\affiliation{Department of Semiconductor Systems Engineering, Korea University, Seoul 02841, Korea}

\author{Se-Hyeok Oh}
\affiliation{Department of Nano-Semiconductor and Engineering, Korea University, Seoul 02841, Korea}

\author{Gyungchoon Go}
\affiliation{Department of Materials Science and Engineering, Korea University, Seoul 02841, Korea}

\author{Jung Hyun Oh}
\affiliation{Department of Materials Science and Engineering, Korea University, Seoul 02841, Korea}

\author{Kyoung-Whan Kim}
\affiliation{Center for Spintronics, Korea Institute of Science and Technology, Seoul 02792, Korea}

\author{Hyun-Woo Lee}
\affiliation{Department of Physics, Pohang University of Science and Technology, Pohang 37673, Korea}

\author{Kyung-Jin Lee}
\email{kj\_lee@korea.ac.kr}
\affiliation{KU-KIST Graduate School of Converging Science and Technology, Korea University, Seoul 02841, Korea}
\affiliation{Department of Materials Science and Engineering, Korea University, Seoul 02841, Korea}

\date{\today}

\begin{abstract}
We numerically compute current-induced spin-transfer torques for antiferromagnetic domain walls, based on a linear response theory in a tight-binding model. We find that, unlike for ferromagnetic domain wall motion, the contribution of adiabatic spin torque to antiferromagnetic domain wall motion is negligible, consistent with previous theories. As a result, the non-adiabatic spin-transfer torque is a main driving torque for antiferromagnetic domain wall motion. Moreover, the non-adiabatic spin-transfer torque for narrower antiferromagnetic domain walls increases more rapidly than that for ferromagnetic domain walls, which is attributed to the enhanced spin mistracking process for antiferromagnetic domain walls.
\end{abstract}


\maketitle
\section{INTRODUCTION}\label{sec:intro}
Antiferromagnetic spintronics has recently attracted considerable interest because of the immunity against external magnetic fields and the potential for high frequency dynamics~\cite{MacDonald, Duine,Tomas}.
Antiferromagnets produce no stray field and do not couple to external magnetic fields because of zero net magnetic moment, which is advantageous for high-density device integration.
Moreover, in contrast to ferromagnets, the resonance frequency of antiferromagnets for the zero wavevector mode is related to the exchange interaction, which results in terahertz magnetic excitations~\cite{Satoh,Kampfrath} and may find use in terahertz spintronic devices~\cite{Cheng2016,Johansen,Khymyn,DKLee}. For domain wall dynamics, it was predicted that spin-orbit torques enable much faster antiferromagnetic domain wall motion than ferromagnetic counterpart~\cite{Ohshba,Gomonay}. This fast domain wall dynamics is caused by the complete decoupling between the domain wall position and domain wall angle because the gyrotropic coupling is proportional to the net spin density~\cite{SKKim,Kabjin,Caretta,Siddiqui,SHJOM}, which is zero in antiferromagnets. 

Antiferromagnets can also be electrically manipulated by conventional spin-transfer torques in the absence of the spin-orbit interaction~\cite{EarlySTT1,EarlySTT2,EarlySTT3,EarlySTT4,Saidaoui,Cheng2014,Xu2008,Swaving2011,Hals2011,Tveten2013,Zhang2015arXiv,Barker2015arXiv,Yuta}.
Previous studies on conventional spin-transfer torques can be classified into two groups. The first group~\cite{EarlySTT1,EarlySTT2,EarlySTT3,EarlySTT4,Saidaoui,Cheng2014,Xu2008} is for spin-valve-like structures in which an antiferromagnet is interfaced with a normal metal and the spin-transfer torque consists of damping-like and field-like components through real and imaginary spin-mixing conductances at the antiferromagnet/normal metal interface~\cite{Cheng2014,Yaro}. The second group~\cite{Xu2008,Swaving2011,Hals2011,Tveten2013,Zhang2015arXiv,Barker2015arXiv,Yuta} is for continuously varying antiferromagnetic spin textures such as domain walls for which the spin-transfer torque consists of adiabatic and non-adiabatic torques. An {\it ab initio} study~\cite{Xu2008} computed the adiabatic spin torque for a domain wall in a system where two ferromagnetic layers are antiferromagnetically coupled in the thickness direction. A microscopic calculation based on the Green's function formulation of Landauer-B{\"u}ttiker transport theory~\cite{Swaving2011} reported that non-equilibrium spin density corresponding to the adiabatic torque for antiferromagnetic spin textures is finite but does not lead to domain wall motion. Phenomenological theories based on spin pumping and Onsager reciprocity~\cite{Hals2011,Tveten2013} predicted that the main driving torque for antiferromagnetic domain wall motion is the non-adiabatic torque. For antiferromagnetic domain walls, therefore, no microscopic computation of both torque components on equal footing has been reported. In order to understand distinct features of adiabatic and non-adiabatic spin-transfer torques acting on antiferromagnetic spin textures, it is important to microscopically compute these two mutually orthogonal torque components.

For ferromagnetic domain walls, a number of theoretical~\cite{Tatara,Zhang,
Thiaville,Yaro2,Stiles,Tatara2,Garate,Gilmore1,Gilmore2,Manchon2,Manchon,Kim2015} and experimental studies~\cite{Hayashi,Moriya,Heyne,Boulle,Eltschka,Burrowes,Sekiguchi,Chauleau,Bisig} have investigated the non-adiabaticity $\beta$ of spin currents, a key parameter of non-adiabatic spin-transfer torque. Different mechanisms of $\beta$ arise depending on the relative length scale of domain wall. One mechanism that is independent of the domain wall length is caused by the spin relaxation~\cite{Zhang,
Thiaville,Yaro2,Garate,Gilmore1,Gilmore2}. This mechanism predicts $\beta/\alpha \approx 1$ where $\alpha$ is the damping parameter, which is related to the spin relaxation in equilibrium. Another mechanism that is independent of the domain wall length is the intrinsic spin torque due to the perturbation of the electronic states when an electric field is applied~\cite{Kim2015}. As domain walls get narrower than the length scale of spin precession around the exchange field, other mechanisms become more dominant. For narrow domain walls, the conduction electron spins are unable to follow a rapid change in the magnetization, i.e. the ballistic spin mistracking, which contributes to the non-adiabaticity~\cite{Stiles,Ohe,Tatara2}. When domain walls are atomically thin, the reflection of conduction electron spins from the domain wall becomes non-negliglible, resulting in the momentum transfer~\cite{Tatara}. A recent experiment~\cite{Okuno} reported a large $\beta/\alpha$ for a domain wall in an antiferromagnetically coupled ferrimagnet, suggesting that mechanisms beyond the spin relaxation may take effect in antiferromagnetic domain walls.

In this paper, we compute non-equilibrium spin density based on a linear response theory in a tight binding model. From the computed non-equilibrium spin density that is defined at each sublattice, we calculate {\it local} and {\it effective} spin-transfer torques, which can be decomposed to adiabatic and non-adiabatic torques. Here {\it local} spin-transfer torque is a torque exerting on a spin moment at each sublattice whereas {\it effective} spin-transfer torque is obtained by integrating local spin-transfer torque over the antiferromagnetic domain wall profile. Therefore, the effective spin-transfer torque is the experimentally measurable quantity. As the leading-order contribution, we find that the effective adiabatic torque is zero for antiferromagnetic domain walls. On the other hand, the effective non-adiabatic torque is large and increases significantly as the domain wall width decreases.

The paper is organized as follows. In Sec. \ref{sec:model}, we present a linear response theory to compute the local spin-transfer torque at sublattices of an antiferromagnetic domain wall. In Sec. \ref{sec:form}, we present a continuum approximation of the equations of motion of an antiferromagnetic domain wall and describe how to obtain the effective spin-transfer torques integrated over the antiferromagnetic domain wall profile from the local spin-transfer torques. In Sec. \ref{sec:result}, we show numerical results of the local and effective spin-transfer torques for antiferromagnetic domain walls and compare the results with ones for ferromagnetic domain walls. Finally, Sec. \ref{sec:conclusion} concludes this work.

\section{\label{sec:model} Microscopic approach to compute spin-transfer torques}

For an antiferromagnet, we consider two sublattices, $A$ and $B$, alternating in the $x$-direction along which the magnetization profile has a texture (i.e., domain wall) and an electric field is applied.
We model a magnetic system with random impurities under an external electric field 
as ${\cal H}=\sum_{\bf k}[{\cal H}^0({\bf k})+V^{\rm imp}+{\cal H}^{\rm ext}({\bf k},t)]$, where $V^{\rm imp}$ is the impurity potential and {\bf k} is a wave vector in the transverse direction, and
\begin{eqnarray}
{\cal H}^0({\bf k}) &\!=\!& \sum_{i}
\left( \begin{array}{cc} {\bf C}_{iA}^\dagger({\bf k}) & {\bf C}_{iB}^\dagger({\bf k}) \end{array} \right)
\left( \begin{array}{cc}
\epsilon({\bf k})\sigma_0\!-\!\Delta_A{\pmb\sigma}\!\cdot\!{\bf m}_{iA} & 0\\
0 & \epsilon({\bf k})\sigma_0\!-\!\Delta_B{\pmb\sigma}\!\cdot\!{\bf m}_{iB} \\
\end{array} \right)
\left( \begin{array}{c} {\bf C}_{iA}({\bf k}) \\ {\bf C}_{iB}({\bf k}) \end{array} \right)
\nonumber\\
&-& t_H \sum_{i} \left[  {\bf C}_{iA}^\dagger({\bf k}) \sigma_0 {\bf C}_{iB}({\bf k}) +  {\bf C}_{iB}^\dagger({\bf k}) \sigma_0 {\bf C}_{i+1,A} ({\bf k})\right]\nonumber\\
&-& t_H \sum_{i} \left[  {\bf C}_{iB}^\dagger({\bf k}) \sigma_0 {\bf C}_{iA}({\bf k}) +  {\bf C}_{i+1,A}^\dagger({\bf k}) \sigma_0 {\bf C}_{i,B}({\bf k}) \right], 
\label{Hm0}
\end{eqnarray}
where $\Delta_\eta$ is an exchange strength at an atomic sublattice ($\eta=A,B$), ${\bf m}_{i\eta}$ is the unit vector along the magnetic moment at a sublattice $\eta$ in the $i$-th cell, $t_H$ is a hopping energy between sublattices,
${\pmb\sigma}$ is the Pauli spin matrix, and
$\sigma_0$ is a $2\times 2$ identity matrix. Because our model assumes alternating sublattices in the $x$-direction, we consider the nearest-neighbor hopping between two sublattices. 
We specify creation operators with ${\bf C}^\dagger_{i\eta}({\bf k})=(C_{i\eta\uparrow}^\dagger({\bf k}),C^\dagger_{i\eta\downarrow}({\bf k}))$
at a sublattice $\eta$ in the $i$-th cell with spins ($\uparrow$ or $\downarrow$) and
annihilation operators with ${\bf C}_{i\eta}({\bf k})=(C_{i\eta\uparrow}({\bf k}),C_{i\eta\downarrow}({\bf k}))^T$, respectively.
We assume that, along the transverse directions (i.e., the $y$- and $z$-directions), the system keeps a periodic structure so that quantum states in these transverse directions are
described by a wave vector ${\bf k}=(k_y, k_z)$ and eigenenergy $\epsilon({\bf k})=-2t_H(\cos k_y d+\cos k_z d)$, where $d$ is the atomic spacing.

On the other hand, ${\cal H}^{\rm ext}({\bf k},t)$ describes the external electric field;
\begin{eqnarray}
{\cal H}^{\rm ext}({\bf k},t) &=& 
-\frac{i|e| t_H d }{\hbar}
 \sum_{i} {\bf {\hat x}}\cdot{\bf A}_i(t) \left[  {\bf C}_{iA}^\dagger({\bf k}) \sigma_0 {\bf C}_{iB}({\bf k}) +  {\bf C}_{iB}^\dagger({\bf k}) \sigma_0 {\bf C}_{i+1,A}({\bf k}) \right]
\nonumber\\
 && +
 \frac{i|e| t_H d }{\hbar}
\sum_{i} {\bf {\hat x}}\cdot{\bf A}_i(t) \left[  {\bf C}_{iB}^\dagger({\bf k}) \sigma_0 {\bf C}_{iA}({\bf k}) +  {\bf C}_{i+1,A}^\dagger({\bf k}) \sigma_0 {\bf C}_{i,B}({\bf k}) \right]
\end{eqnarray}
where ${\bf A}_i(t) = -{\bf {\hat x}}{\cal E} \sin\omega_p t/\omega_p$ is a vector potential with the electric field ${\cal E}$ in the $x$ direction. For a DC case, we set  a frequency $\omega_p$ to be zero at the final stage of calculation.

By integrating out electronic degrees of freedom $\{ {\bf C}^\dagger, {\bf C}\}$ on the Keldysh contour,
a local spin torque at a site $i\eta$ is given by~\cite{Balaz}
\begin{eqnarray}
{\pmb \tau}_{i\eta}= \Delta_\eta {\bf m}_{i\eta}(t) \times \delta{\bf s}(t)_{i\eta} = \Delta_\eta {\bf m}_{i\eta}(t) \times \sum_{\bf k}{\rm Tr}_{\rm spin}\{-i\hbar {\bf G}^<({\bf k};t) {\pmb\sigma}\}_{i\eta,i\eta}
.
\label{taudetail}
\end{eqnarray}
Here, ${\bf G}^<({\bf k};t)$ is a lesser Green function of the full Hamiltonian, ${\bf G}({\bf k},t)=[i\hbar\partial_t-{\cal H}^0({\bf k})-V^{\rm imp}-{\cal H}^{\rm ext}({\bf k},t)]^{-1}$,
and the non-equilibrium spin density is determined by a linear part of a full Green function on a vector potential ${\bf A}(t)$ as, 
$\delta{\bf s}_{i\eta}(t)= \sum_{\bf k}{\rm Tr}_{\rm spin}\{-i\hbar ({\cal G} H^{\rm ext} {\cal G})^<({\bf k};t) {\pmb\sigma}\}$ where
${\cal G}=[E-{\cal H}^0-V^{\rm imp}]^{-1}$ is the unperturbed Green function averaged over impurities.

\subsection{ Linear response approximation}

We treat the external electric field perturbatively because the energy change between adjacent atoms, $|e| {\cal E}d$, is much smaller than the hopping energy, $t_H$. 
Along this scheme, the lesser Green function of the system is written as~\cite{Lake}
\begin{eqnarray}
{\bf G}^<_{i\eta,i'\eta'}({\bf k};t) = -i\hbar\langle {\bf C}^\dagger_{i'\eta'}({\bf k},t) {\bf C}_{i\eta}({\bf k},t)\rangle
= -\frac{1}{i\hbar} \left[ \rho^{(0)}({\bf k})+ \rho^{(1)}({\bf k},t)+\cdots \right]_{i\eta,i'\eta'}.
\end{eqnarray}
Here each density matrices $\rho^{(n)}$ denote the orders of ${\cal H}^{\rm ext}$ and are defined by,
\begin{eqnarray}
\rho^{(0)}({\bf k}) &=& \frac{i}{2\pi} \int dE  f_{\rm o}(E) {\cal G}^C({\bf k},E),
\nonumber\\
\rho^{(1)}({\bf k},t) &=& \frac{i}{2\pi} \int dE d\omega e^{-i\omega t} f_{\rm o}(E)\nonumber\\
&&\times [{\cal G}^R({\bf k},E\!+\!\hbar\omega) U (\omega) {\cal G}^C({\bf k},E) \!+\!{\cal G}^C({\bf k},E) U (\omega) {\cal G}^A({\bf k},E\!-\!\hbar\omega) ],
\label{Grho}
\end{eqnarray}
where $f_{\rm o}$ is the Fermi-Dirac distribution function,  
${\cal G}^{R,A}$ are retarded and advanced Green functions averaged over impurities with
${\cal G}^C = {\cal G}^R-{\cal G}^A$, and $ U (\omega) = 1/(2\pi i)  \int dt H^{\rm ext}(t) e^{i\omega t}$ is Fourier component of ${\cal H}^{\rm ext}(t)$ for time.
$\rho^{(0)}({\bf k})$ is a density matrix in equilibrium whereas
$\rho^{(1)}({\bf k},t)$ is the non-equilibrium one modified by the external electric field.
Then, within the linear response approximation we set the non-equilibrium spin density as
\begin{eqnarray}
\delta{\bf s}_{i\eta}(t) = \sum_{{\bf k}}{\rm Tr}_{\rm spin} \{ {\pmb \sigma}\rho_{i\eta, i\eta}^{(1)}({\bf k},t) \}.
\label{deltaS}
\end{eqnarray}

\subsection{ Calculation of Green functions }

When a domain wall is present in a magnetic lattice without impurities, the retarded $({\bf g}^R)$ and advanced $({\bf g}^A)$ Green functions associated with ${\cal H}^0$
are given by
\begin{eqnarray}
{\bf g}^R({\bf k},E) &=& [ (E+i\delta){\bf 1}-{\cal H}^0({\bf k})]^{-1},
\nonumber\\
{\bf g}^A({\bf k},E) &=& [ (E-i\delta){\bf 1}-{\cal H}^0({\bf k})]^{-1}= [{\bf g}^R({\bf k},E)]^\dagger,
\end{eqnarray}
with $\delta$ a positive infinitesimal.
Here, ${\cal H}^0({\bf k})$ is the unperturbed Hamiltonian with a transverse state ${\bf k}$ and has an infinite dimension because the cell index $i$ runs $-\infty$
to $\infty$.
However, its inverse matrix is readily obtained because ${\cal H}^0({\bf k})$ is a $4\times 4$ block-tridiagonal matrix
as seen from the Hamiltonian of Eq.~(\ref{Hm0}).

Now let us confine our attention to the region extended over the domain wall. 
When the electric field is applied, electron densities are modified due to the hopping interaction given by ${\cal H}^{\rm ext}$. Namely, $\delta{\bf s}_i$ is determined by contributions from
various neighboring cell $i'$ as indicated from Eq. (\ref{Grho}).
A weight of its contribution is determined by the Green function between $i$ and $i'$ and
is approximately given by ${\bf g}^R_{ii'}\propto e^{-\zeta|i-i'|}$ where a decay constant $\zeta$ is proportional
to the level broadening $\delta$~\cite{Datta1}.
Therefore, as a level broadening becomes larger, the non-equilibrium spin density is contributed from nearer atomic sublattices. 
By assuming a large level broadening due to impurity scattering discussed in the next section,
we consider a finite region but still large enough to compass the domain wall, say the number of $N_{\rm cell}$ cells.
We truncate out the remaining regions ranging over $-\infty<i<0$ and $N_{\rm cell}\leq i<\infty$, whose effects are incorporated into the Green functions with
self-energies $\Sigma^{L,R}$~\cite{Datta1,Lake}. Therefore, the Green functions for the interested region that includes the domain wall are given by
\begin{eqnarray}
{\bf g}^R({\bf k},E) &=& [ (E+i\delta){\bf 1}-H^0({\bf k})-\Sigma^L(E)-\Sigma^R(E)]^{-1},
\nonumber\\
{\bf g}^A({\bf k},E) &=& [{\bf g}^R({\bf k},E)]^\dagger.
\end{eqnarray}
As a result, ${\bf g}^{R,A}({\bf k},E)$ have a dimension of $4N_{\rm cell}\times 4N_{\rm cell}$ that are tractable numerically.

\subsection{ Impurity scattering }

We average the Green functions over impurity configurations~\cite{Kamenev,Ohimp}.
The corresponding self-energy in the coordinate representation is given by 
\begin{eqnarray}
\Sigma^{\rm imp}({\bf r},{\bf r}';E) = {\cal G}^R({\bf r},{\bf r}';E)\langle V^{\rm imp}({\bf r}),V^{\rm imp}({\bf r}')\rangle
\label{SigmaImp0}
\end{eqnarray}
where $\langle V^{\rm imp}({\bf r}),V^{\rm imp}({\bf r}')\rangle$ is the correlation function of impurities and ${\cal G}^R$ is a retarded
Green function averaged over impurities.
For random and short-ranged impurities with a screening length $1/k_s$,  the correlation function is found to be proportional to~\cite{Ohimp}
\begin{equation}
\langle V^{\rm imp}({\bf r})V^{\rm imp}({\bf r}^\prime) \rangle \propto
 \frac{ 1 }{k_s}e^{-k_s\mid {\bf r}-{\bf r}^\prime\mid},
\label{Ucorr}
\end{equation}
indicating that $\Sigma^{\rm imp}$ is also a short-ranged function.
Assuming such short-ranged impurities we set
\begin{equation}
\Sigma_{ii'}^{\rm imp}(E) = \delta_{ii'}\frac{V_0^2}{4\pi^2}\sum_{\bf k}{\cal G}_{ii}^R({\bf k},E)
\label{SigmaIMP}
\end{equation}
where a summation over ${\bf k}$ means that the impurity self-energy is also local over the transverse directions.
Thus, by including the self-energy from impurity, the Green function forms the Dyson equation; 
\begin{equation}
\sum_{i_1}\Big[ [{\bf g}^R({\bf k},E)]_{ii_1}^{-1}-\Sigma_{ii_1}^{\rm imp}(E) \Big] {\cal G}^R_{i_1 i'}({\bf k},E) = {\bf 1}^{4\times4}\delta_{ii'}
\end{equation}
with a $4\times4$ identity matrix ${\bf 1}^{4\times4}$.
A solution of the equation is not trivial due to a self-consistency of ${\cal G}^R$ and
requires a large computation burden.
For simplicity, because the modification of electronic structure by external fields occurs mainly near the chemical potential,
we take into account the self-energy at the chemical potential $\mu$ in a whole energy as \[\Sigma^{\rm imp}(E)\approx \Sigma^{\rm imp}(E)\Big|_{E=\mu},\] namely the self-energy independent of energy.
We solve the Dyson equation self-consistently to obtain $\Sigma^{\rm imp}$ of Eq. (\ref{SigmaIMP}),
and obtain the non-equilibrium spin density by using Eqs. (\ref{Grho}) and ({\ref{deltaS}). This locally defined non-equilibrium spin density gives a local spin torque through Eq.~(\ref{taudetail}).

\section{\label{sec:form}Effective spin-transfer torques acting on an antiferromagnetic domain wall}
To investigate the role of spin-transfer torques in domain wall motion, one has to find out the effective spin-transfer torques, which are obtained by integrating local spin-transfer torques over the domain wall profile. This section presents the equations of motion and associated effective spin-transfer torques acting on an antiferromagnetic domain wall as follows.

Local magnetic moment at each sublattice in the $i^{th}$ unit cell is assumed to be
${\bf m}_{i\eta}={\rm sign}_\eta (\cos \phi_{i\eta} \sin \theta_{i\eta}, \sin \phi_{i\eta} \sin \theta_{i\eta}, \cos \theta_{i\eta})$ with $\eta=A,B$.
Here, $\theta_{i\eta}$ and $\phi_{i\eta}$ are polar and azimuthal angles at a sublattice $\eta$ in the $i^{th}$ unit cell.
We introduce a prefactor ${\rm sign}_\eta$ to describe antiferromagnet or ferromagnet (i.e., ${\rm sign}_A=1$ and ${\rm sign}_B=-1$ for antiferromagnet and ${\rm sign}_A={\rm sign}_B=1$ for ferromagnet). We adopt the Walker's ansatz for a domain wall profile~\cite{Landau} as, 
\begin{equation}
\theta_{i\eta}=2 \tan^{-1}\{\exp[(X-x_{i\eta})/ \lambda_{\rm DW}]\},~~~~~~~~\phi_{i\eta}=\frac{\pi}{2}, 
\label{Blochdomain}
\end{equation}
where $X$ and $\lambda_{\rm DW}$ are the position and the width of the domain wall, respectively. 
For antiferromagnetic domain walls (see Fig.~\ref{Fig1} for a domain wall profile), we introduce the total and staggered magnetic moments as
\begin{equation}
{\bf M}_{i} \equiv \frac{1}{2}({\bf m}^{\rm AFM}_{iA}+{\bf m}^{\rm AFM}_{iB}),
~~~~{\bf n}_{i} \equiv \frac{1}{2}({\bf m}^{\rm AFM}_{iA}-{\bf m}^{\rm AFM}_{iB}).
\end{equation}

The equations of motion for antiferromagnetic domain walls are obtained with the second-order expansion of small parameters ($\partial /\partial x$, $\partial /\partial t$, $\mathbf{M}$, and spin-torque terms). The free energy ${\cal U}$ of the system is written as~\cite{Tveten2016, SKKim, Ohshba}
\begin{equation}
{\cal U} = \int dx\left[\frac{a}{2} |{\bf M}|^{2}+\frac{A}{2}\left(\frac{\partial{\bf n}}{\partial x}\right)^{2}
+L{\bf M}\cdot\frac{\partial{\bf n}}{\partial x}\right],
\label{freeE}
\end{equation}
where $a$ ($A$) is homogeneous (inhomogeneous) exchange parameter and $L$ is a parity-breaking exchange strength~\cite{Papanicolaou,Tveten2016}. The Euler-Lagrange equation with respect to ${\bf M}$ and ${\bf n}$ is given by
\begin{equation}\label{Lag1}
\frac{\partial {\cal L}}{\partial\mathbf{M}(\mathbf{n})}-\frac{\partial}{\partial t}
\left(\frac{\partial{\cal L}}{\partial\dot{\mathbf{M}}\left(\dot{\mathbf{n}}\right)}\right)
=\frac{\partial{\cal R}}{\partial\dot{\mathbf{M}}\left(\dot{\mathbf{n}}\right)},
\end{equation}
where the Lagrangian density ${\cal L}$ and the Rayleigh function ${\cal R}$ are respectively given by~\cite{SKKim, Hals2011, Andreev, Chiolero, Ivanov, S.-H.Oh}
\begin{eqnarray}
\label{Lag2}
{\cal L}&=&s\dot{\mathbf{n}}\cdot(\mathbf{n}\times\mathbf{M})-{\cal U},
\nonumber\\
{\cal R}&=&s\alpha\dot{\mathbf{n}}^{2},
\end{eqnarray}
with $\alpha$ the Gilbert damping parameter, and $s(\equiv M_{s}/\gamma)$ the averaged angular momentum for two sublattices, where $M_{s}$ is the saturation magnetization, and $\gamma$ is the gyromagnetic ratio.

From Eqs. (\ref{freeE})-(\ref{Lag2}), we obtain the equations of motion for $\mathbf{n}$ and $\mathbf{M}$  as
\begin{eqnarray}\label{eqmnM}
\dot{\mathbf{n}}&=&\frac{1}{s}\mathbf{f}_{\mathbf{M}}\times\mathbf{n}+{\pmb \tau}_{\mathbf{n}},\nonumber\\
\dot{\mathbf{M}}&=&\frac{1}{s}\mathbf{f}_{\mathbf{n}}\times\mathbf{n}-2\alpha\dot{\mathbf{n}}\times\mathbf{n}+{\pmb \tau}_{\mathbf{M}},
\end{eqnarray}
where $\mathbf{f}_{\mathbf{M}(\mathbf{n})}$ is an effective field on $\mathbf{M}(\mathbf{n})$,
$\mathbf{f}_{\mathbf{M}}\equiv -a\mathbf{M}-L\tfrac{\partial\mathbf{n}}{\partial x}$,
 $\mathbf{f}_{\mathbf{n}}\equiv A\tfrac{\partial^{2}\mathbf{n}}{\partial x^{2}}+L\tfrac{\partial\mathbf{M}}{\partial x}$, and ${\pmb \tau}_{\bf M}$ and ${\pmb \tau}_{\bf n}$ are respectively spin-transfer torques acting on ${\mathbf{M}}$ and ${\mathbf{n}}$, which we define below.

In ferromagnets, the torques can be decomposed to adiabatic and non-adiabatic components as~\cite{Zhang, Tatara, Thiaville}
\begin{equation}\label{FMSTT}
{\pmb \tau}_{\rm FM} =  \tau_{\rm FM}^{\rm a}\frac{\partial {\bf m}}{ \partial x}
-\tau_{\rm FM}^{\rm na} {\bf m}\times\frac{\partial {\bf m}}{ \partial x},
\end{equation}
with $\tau^{\rm a}$ and $\tau^{\rm na}$, the magnitudes of adiabatic and non-adiabatic torques. Likewise, the spin-transfer torques in each sublattice of an antiferromagnet can be written as
\begin{eqnarray}\label{AFSTT1}
{\pmb \tau}_{A} &=& \tau_A^{\rm a}\frac{\partial {\bf m}^{\rm AFM}_A}{\partial x}
-\tau_A^{\rm na} \left(\mathbf{m}^{\rm AFM}_{\rm A}\times\frac{\partial\mathbf{m}^{\rm AFM}_{A}}{\partial x}\right),
\nonumber\\
{\pmb \tau}_{B} &=& \tau_B^{\rm a}\frac{\partial\mathbf{m}_B^{{\rm AFM}}}{\partial x}
-\tau_{B}^{\rm na} \left(\mathbf{m}^{{\rm AFM}}_B\times\frac{\partial\mathbf{m}^{{\rm AFM}}_{B}}{\partial x}\right).
\end{eqnarray}

Using the total and staggered magnetizations, Eq.~(\ref{AFSTT1}) with the second-order expansion of small parameters becomes~\cite{Hals2011,SKKim}
\begin{eqnarray} \label{AFSTT2}
{\pmb \tau}_{\mathbf{n}} &=& \tau_{\rm AFM}^{\rm a}\frac{\partial \mathbf{n}}{\partial x}, \mspace{18mu}
\nonumber\\
{\pmb \tau}_{\mathbf{M}} &=&-\tau_{\rm AFM}^{\rm na}\mathbf{n}\times\frac{\partial \mathbf{n}}{\partial x},
\end{eqnarray}
where 
\begin{equation}
\tau^{\rm a}_{\rm AFM} = \frac{\tau^{\rm a}_{A}+\tau^{\rm a}_{B}}{2},~~~
\tau^{\rm na}_{\rm AFM}=\frac{\tau^{\rm na}_{A}+\tau^{\rm na}_{B}}{2}.
\end{equation}
We note that in Eq.~(\ref{AFSTT2}), ${\pmb \tau}_{\bf{n}}$ and ${\pmb \tau}_{\bf{M}}$ are obtained from ${\pmb \tau}_{A}$ and ${\pmb \tau}_{B}$, which are computed using the linear response theory described in the above section. One can then obtain $\tau^{\rm a}_{\rm AFM}$ and $\tau^{\rm na}_{\rm AFM}$ from Eq. ~(\ref{AFSTT2}), which are local quantities defined in a unit cell.

\begin{figure}[ttbp]
\begin{center}
\includegraphics[width=\columnwidth]{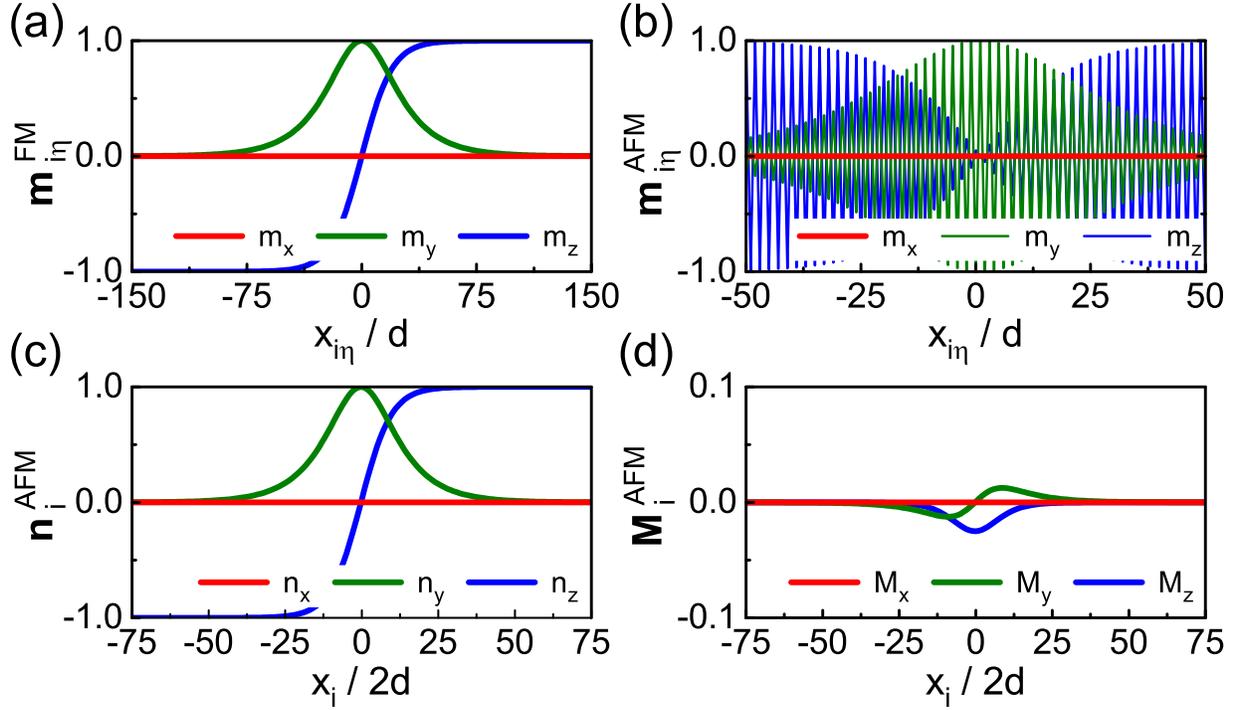}
\caption{(color online) Domain wall profiles of (a) ferromagnetic and (b) antiferromagnetic domain walls, where $d$ is the atomic spacing and the domain wall width $\lambda_{\rm DW}$ is 20$d$.
For the antiferromagnetic domain wall (b), the components of the staggered ${\bf n}$ and total ${\bf M}$ magnetic moments are shown in (c) and (d), respectively. 
}
\label{Fig1}
\end{center}
\end{figure}


To obtain effective spin-transfer torques, we integrate the local spin torques over the domain wall profile as follows.
Using the collective coordinate approach with respect to the domain wall position $X$ and the domain wall angle $\phi$, the equations of motion of an antiferromagnetic domain wall are readily obtained as~\cite{Ohshba, Okuno, Tretiakov}
\begin{eqnarray}\label{eqmXP1}
\rho\ddot{X}+2\alpha s \dot{X} &= &-s\tilde{c}_{J}^{\rm AFM},\nonumber\\
\rho\ddot{\phi}+2\alpha s\dot{\phi} &= &0,
\end{eqnarray}
where $\rho\equiv s^{2}/a$, and  $\tilde{c}_{J}^{\rm AFM}$ is the effective non-adiabatic spin-transfer torque integrated over the antiferromagnetic domain wall profile, given as
\begin{equation}\label{AFNA}
\tilde{c}_{J}^{\rm AFM} =-\frac{\lambda_{\rm DM}}{2}\int^{\infty}_{-\infty}dx \left[\tau^{\rm na}_{\rm AFM}\frac{\partial\mathbf{n}}{\partial x}\cdot\frac{\partial\mathbf{n}}{\partial X}\right].
\end{equation}
By the same way, one can calculate the effective adiabatic spin-transfer torque, $\tilde{b}_{J}^{\rm AFM}$, integrated over the antiferromagnetic domain wall profile, as
\begin{equation}\label{AFA}
\tilde{b}_{J}^{\rm AFM} =\frac{L}{2a}\int^{\infty}_{-\infty}dx\left[\tau^{\rm a}_{\rm AFM}\left(\mathbf{n}\times\frac{\partial^{2}\mathbf{n}}{\partial x^{2}}\right)\cdot\frac{\partial\mathbf{n}}{\partial\phi}\right].
\end{equation}
However, this adiabatic torque contribution, which is the third order of small parameters, is absent in Eq.~(\ref{eqmXP1}) because Eq.~(\ref{eqmXP1}) is obtained by expanding up to the second order. The adiabatic torque contribution appears only when the higher order terms are considered as in Ref.~[\onlinecite{Swaving2011}]. This means that the adiabatic torque contribution to the antiferromagnetic domain wall motion is much weaker than the non-adiabatic torque contribution. Therefore, as a leading-order contribution, the antiferromagnetic domain wall velocity in the steady state (i.e., $\ddot{X}=0$) is determined by the effective non-adiabatic torque, given as
\begin{equation}
v_{\rm DW}=-\frac{\tilde{c}_{J}^{\rm AFM}}{2\alpha}.
\end{equation}

On the other hand, the equations of motion of ferromagnetic domain wall are given as~\cite{Tatara, Thiaville}
\begin{equation}\label{FMmotion1}
\dot{\phi}+\frac{\alpha}{\lambda_{\rm DW}}\dot{X}=-\frac{\tilde{c}_{J}^{\rm FM}}{\lambda_{\rm DW}}, \mspace{18mu}
-\frac{1}{\lambda_{\rm DW}}\dot{X}+\alpha\dot{\phi}=\frac{\tilde{b}_{J}^{\rm FM}}{\lambda_{\rm DW}},
\end{equation}
where~\cite{KJLee}
\begin{eqnarray}\label{FMmotion2}
\tilde{c}_{J}^{\rm FM} &=& -\frac{\lambda_{\rm DW}}{2} \int^{\infty}_{-\infty} dx 
\left[\tau^{\rm na}_{\rm FM}\frac{\partial\mathbf{m}}{\partial x}\cdot\frac{\partial\mathbf{m}}{\partial X}\right].\nonumber\\
\tilde{b}_{J}^{\rm FM} &=& \frac{1}{2}\int^{\infty}_{-\infty}dx\left[\tau^{\rm a}_{\rm FM}
\left(\mathbf{m}\times\frac{\partial \mathbf{m}}{\partial x}\right)\cdot\frac{\partial\mathbf{m}}{\partial \phi}\right].
\end{eqnarray}
As well known, for ferromagnetic domain walls, both adiabatic and non-adiabatic contributions appear in the equations of motion at the same order, in contrast to the case of antiferromagnetic domain wall motion.

In the next section, we present numerical results of local and effective spin-transfer torques for ferromagnetic and antiferromagnetic domain walls.


\begin{figure}[ttbp]
\begin{center}
\includegraphics[width=0.8\columnwidth]{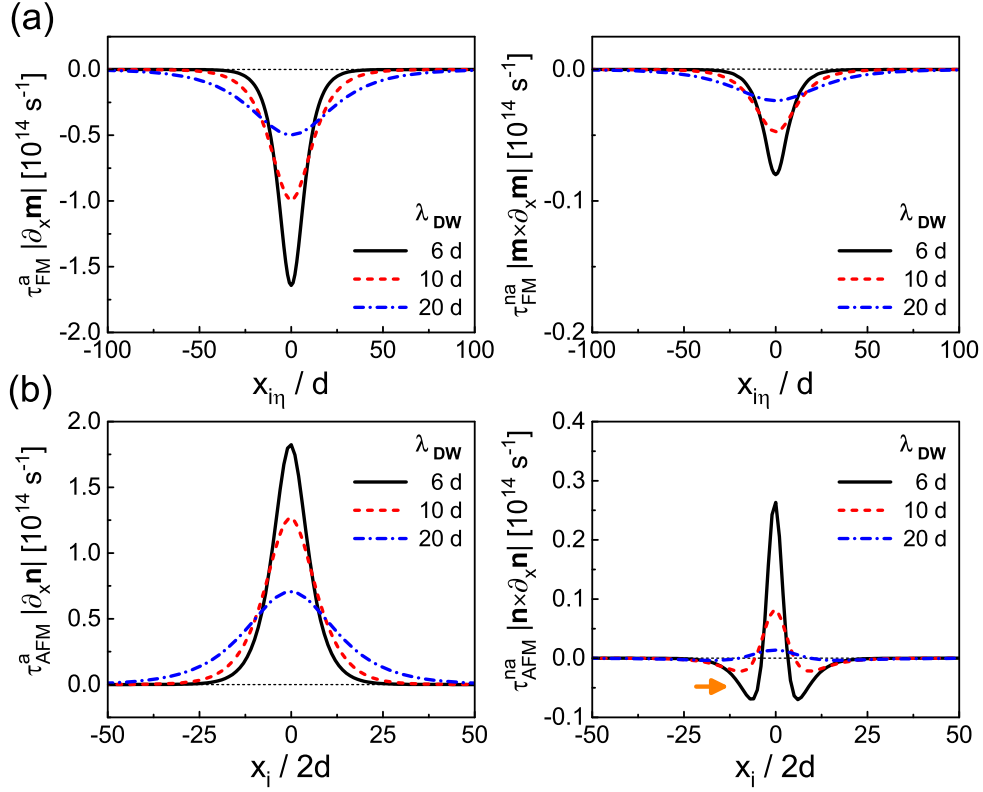}
\caption{(color online) Computed results of local spin-transfer torques. (a) The adiabatic $\tau^{\rm a}$ (left panel) and non-adiabatic $\tau^{\rm na}$
(right panel) torques for ferromagnetic domain walls. (b) The adiabatic $\tau^{\rm a}$ (left panel) and non-adiabatic $\tau^{\rm na}$
(right panel) torques for antiferromagnetic domain walls. In (a) and (b), we compare results for several domain wall widths, $\lambda_{\rm DW}=6d, 10d$ and $20d$. The orange arrow in the right panel of (b) shows negative local non-adiabatic torques for an antiferromagnetic domain wall. 
}
\label{Fig2}
\end{center}
\end{figure}

\section{\label{sec:result} Results and discussion }
\subsection{\label{subsec:STT} Local spin-transfer torques }
 For numerical computation, we choose the number of unit cell $N_{\rm cell}=600$, the atomic spacing $d=0.27$ nm, the hopping parameter $t_{H}=1 ~{\rm eV}$, and the number of ${\bf k}$ point in the transverse direction is $150\times 150$, which guarantees converged results.
We use the Fermi energy $E_{\rm F}=0.0 ~{\rm eV}$, the exchange strength $\Delta=1.0~{\rm eV}$, and the impurity scattering energy parameter $V_{0}=3.5~ {\rm eV}$, unless specified.

 
We calculate the non-equilibrium spin density at each atomic site $i\eta$ using Eq.~(\ref{deltaS}) and decompose
it into local adiabatic and non-adiabatic torques using Eqs.~(\ref{FMSTT}) and (\ref{AFSTT2}).
Calculated local spin-transfer torques are shown in Fig.~\ref{Fig2} for (a) ferromagnetic and (b) antiferromagnetic domain walls. An interesting observation is that the local non-adiabatic torque for a relatively narrow antiferromagnetic domain wall changes its sign near the domain wall center ($x_i=0$), indicated by an orange arrow. This negative local torque originates from spatial oscillation of non-equilibrium spin density near the domain wall, which results from the spin mistracking process~\cite{Stiles,KJLee}. We note that for the same domain wall width ($\lambda_{\rm DW}=6d$), the local torques for ferromagnetic domain walls do not show such sign change, suggesting that the spin mistracking is more pronounced for antiferromagnetic domain walls than for ferromagnetic domain walls. This enhanced spin mistracking for antiferromagnetic domain walls may be understood as follow.  
According to Ref.~[\onlinecite{Stiles}], for ferromagnetic domain walls, the non-adiabaticity due to the spin mistracking process is proportional to $\exp(-\kappa \lambda_{\rm DW}/\zeta)$ where $\kappa$ is a constant, $\zeta=E_F/(\Delta k_F)$, and $k_F$ is the Fermi wave vector. Therefore, the non-adiabaticity increases exponentially with decreasing the exchange interaction $\Delta$. In antiferromagnets, the effective exchange interaction averaged over two sublattices is zero. As a result, it is expected that the characteristic length scale $\zeta$ is very long. We note that the large non-adiabaticity or long characteristic length scale of transverse spin currents were recently reported in experiments using antiferromagnetically coupled ferrimagnets~\cite{Okuno, Yang}.

Another interesting observation is that the local adiabatic torque is sizable for both ferromagnetic [left panel of Fig.~\ref{Fig2}(a)] and antiferromagnetic [left panel of Fig.~\ref{Fig2}(b)] domain walls. We will discuss the relation between this non-zero local adiabatic torque and effective adiabatic torque for antiferromagnetic domain walls in the next section. Finally, in Fig.~\ref{Fig2}, it is observed that the signs of the torque are different for ferromagnetic and antiferromagnetic domain walls. However, this sign difference is found to depend on the parameters (not shown), which may depend on band details~\cite{Garate}.

\subsection{\label{subsec:result2}Effective spin-transfer torques for antiferromagnetic domain walls}
\begin{figure*}[ht]
\begin{center}
\includegraphics[width=13cm]{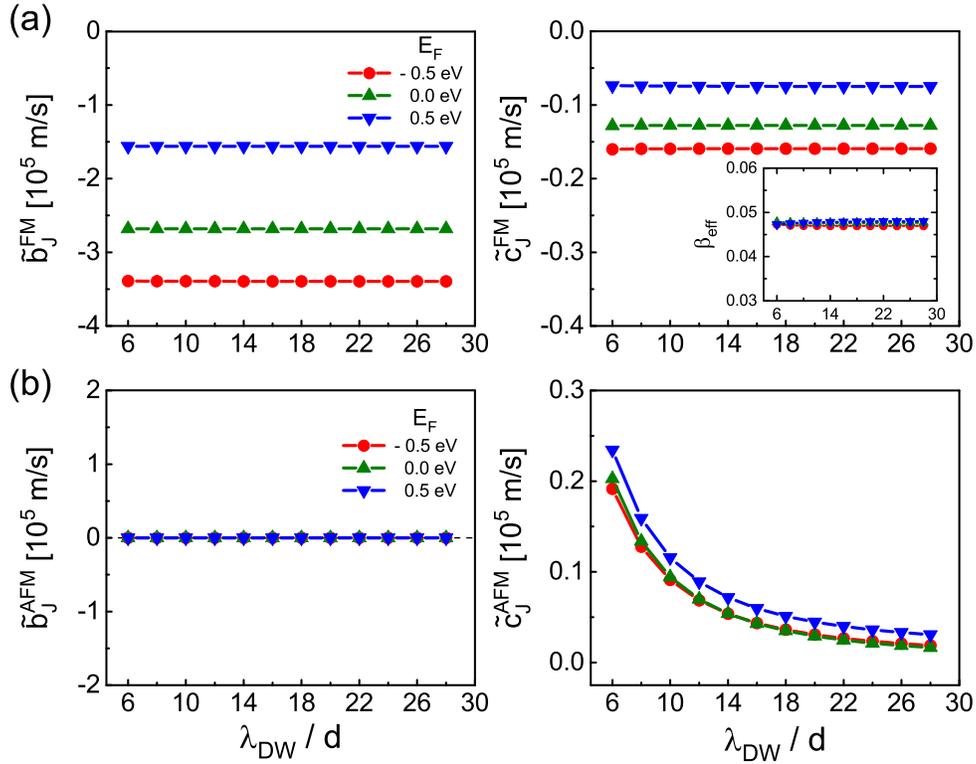}
\caption{ (color online) 
In (a), the effective adiabatic torque ${\tilde {\rm b}}_{\rm J}$ (left panel) and the effective non-adiabatic torque ${\tilde{\rm c}}_J$ (right panel) for ferromagnetic domain walls are plotted as a reduced domain wall widths $\lambda_{DW}/d$.
We compare results for several Fermi energies, $E_F= -0.5$ eV, $0.0$ eV, and $0.5$ eV.
In (b), under the same condition we examine the effective adiabatic torque (left panel) and the effective non-adiabatic torque (right panel) for antiferromagnetic domain walls.
The inset in the right panel of (a) shows the effective non-adiabaticity $\beta_{\rm eff}$ in the ferromagnet case.
\label{Fig3}
}
\end{center}
\end{figure*}

In this section, we discuss effective adiabatic spin-transfer torques ($\tilde{b}_{J}^{\rm FM}$ and $\tilde{b}_{J}^{\rm AFM}$) and effective non-adiabatic spin-transfer torques ( $\tilde{c}_{J}^{\rm FM}$ and $\tilde{c}_{J}^{\rm AFM}$), which are calculated by integrating the local torques over the domain wall profile [see Eqs.~(\ref{AFNA}), (\ref{AFA}), and (\ref{FMmotion2})]. 

For ferromagnetic domain walls [Fig.~\ref{Fig3}(a)], the effective adiabatic ($\tilde{b}_{J}^{\rm FM}$; left panel) and non-adiabatic ($\tilde{c}_{J}^{\rm FM}$; right panel) torques are almost constant regardless of the domain wall width ranging from $6d$ to $28d$. Even with a variation of $E_F$, this insensitivity to the domain wall width is maintained. Since both $\tilde{b}_{J}^{\rm FM}$ and $\tilde{c}_{J}^{\rm FM}$ are finite, one can define the effective non-adiabaticity $\beta_{\rm eff} (\equiv \tilde{c}_{J}^{\rm FM}/\tilde{b}_{J}^{\rm FM}$), which is almost a constant of the order of 0.05 in our model, consistent with previous works~\cite{Eltschka, Burrowes, Sekiguchi, Chauleau}.
In contrast, the effective torques for antiferromagnetic domain walls show two distinct features in comparison to those for ferromagentic domain walls. First, the effective adiabatic torque [$\tilde{b}_{J}^{\rm AFM}$; left panel of Fig.~\ref{Fig3}(b)] is almost zero regardless of the Fermi energy and domain wall width. Given that the local adiabatic torque for antiferromagnetic domain walls is finite [left panel of Fig.~\ref{Fig2}(b)], this nearly zero effective adiabatic torque results from the symmetry of $ \left({\bf n} \times\frac{\partial^2{\bf n}}{\partial x^2}\right)\cdot\frac{\partial{\bf n}}{\partial \phi}$, which is zero when integrating over a whole domain wall profile [see the integral of Eq.~(\ref{AFA})]. It also supports that the adiabatic torque contribution to the antiferromagnetic domain wall motion is almost absent. 

Second, the effective non-adiabatic torque [$\tilde{c}_{J}^{\rm AFM}$; right panel of Fig.~\ref{Fig3}(b)] increases rapidly with decreasing the domain wall width, which is consistent with that expected for the spin mistracking process. To further validate the spin mistracking process as a main origin of the enhanced $\tilde{c}_{J}^{\rm AFM}$ for a narrower wall, we compute $\tilde{c}_{J}^{\rm FM}$ and $\tilde{c}_{J}^{\rm AFM}$ with varying the exchange parameter $\Delta$ (Fig.~\ref{Fig4}). We find that $\tilde{c}_{J}^{\rm AFM}$ increases more rapidly than $\tilde{c}_{J}^{\rm FM}$. These results support that the spin mistracking process is responsible for the enhanced $\tilde{c}_{J}^{\rm AFM}$ for a narrower wall, especially in antiferromagnets.

\begin{figure}[ttbp]
\begin{center}
\includegraphics{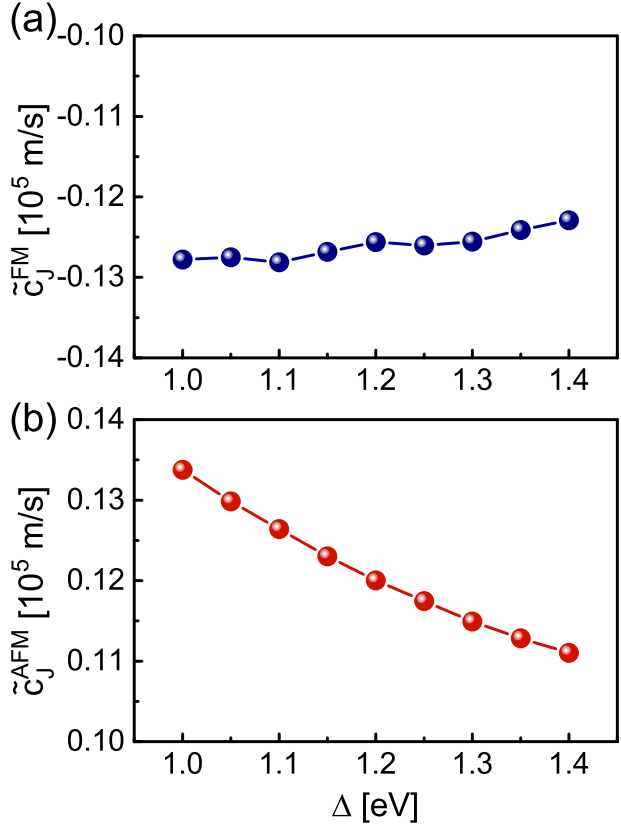}
 \caption {\label{Fig4} (color online) The effective non-adiabatic torque $\tilde{c}_{J}$ with various exchange parameter $\Delta$ in (a) ferromagnets and (b) antiferromagnets at domain wall width $\lambda_{\rm DW}=8d$. 
 }
\end{center}
\end{figure}

\section{\label{sec:conclusion}conclusion}
In this paper, we numerically compute the adiabatic and non-adiabatic spin-transfer torques for antiferromangetic domain walls. We find that the effective adiabatic torque in antiferromagnetic domain walls is almost zero, which means that the adiabatic torque does not affect dynamics of antiferromagnetic domain walls. This negligible contribution of the adiabatic spin torque to antiferromagnetic domain wall motion is consistent with previous theories~\cite{Hals2011,Tveten2013} based on spin pumping and Onsager reciprocity. It is also consistent with a recent experiment~\cite{Okuno} showing that the adiabatic torque contribution on the velocity of ferrimagnetic domain wall is proportional to the equilibrium net spin density $\delta_{s}$ and is thus almost zero near the angular momentum compensation temperature $T_{\rm A}$.

We also find that the effective non-adiabatic torque for antiferromagnetic domain walls can be sizable and increases more rapidly with decreasing the domain wall width in comparison to that for ferromagnetic domain walls. Our result supports that the rapid increase of non-adiabatic torque for antiferromagnetic domain walls is caused by the spin mistracking process, which is more pronounced in antiferromagnets than in ferromagnets. 

As a final remark, given that the effective adiabatic torque $\tilde{b}_{J}^{\rm AFM}$ is almost zero while the effective non-adiabatic torque $\tilde{c}_{J}^{\rm AFM}$ is finite, it is unphysical to define the non-adiabaticity ($\beta=\tilde{c}_{J}^{\rm AFM}/\tilde{b}_{J}^{\rm AFM}$) for antiferromagnetic domain walls. For the same reason, the question about whether or not $\beta$ is close to the damping constant $\alpha$, which has been a long-standing debate for ferromagnetic domain walls~\cite{Kohno2006, Yaro3, Duine2007, Garate, Boulle2011}, is not justified for antiferromagnetic domain walls.


\begin{acknowledgments}
This work was supported by the National Research Foundation of Korea (NRF) (NRF-2015M3D1A1070465, NRF-2017R1A2B2006119) and by the Korea Institute of Science and Technology (KIST) Institutional Program (project no. 2V05750, 2E29410). G.G. was supported by NRF-2019R1I1A1A01063594. H.-W.L. was supported by NRF-2018R1A5A6075964.
\end{acknowledgments}



\begin{thebibliography}{99}

\bibitem{MacDonald} A. H. MacDonald and M. Tsoi, Phil. Trans. R. Soc. A {\bf 369}, 3098 (2011).
\bibitem{Duine} R. Duine, Nat. Mater. {\bf 10}, 344 (2011).
\bibitem{Tomas} T. Jungwirth, X. Marti, P. Wadley, and J. Wunderlich, Nat. Nanotechnol. {\bf 11}, 231 (2016).

%
\bibitem{Satoh} T. Satoh, S.-J. Cho, R. Iida, T. Shimura, K. Kuroda, H. Ueda, Y. Ueda, B. A. Ivanov, F. Nori, and M. Fiebig, Phys. Rev. Lett. {\bf 105}, 077402 (2010).
%
\bibitem{Kampfrath} T. Kampfrath, A. Sell, G. Klatt, A. Pashkin, S. Mahrlein, T. Dekorsy, M. Wolf, M. Fiebig, A. Leitenstorfer, and R. Huber, Nat. Photonics {\bf 5}, 31 (2011).
%
\bibitem{Cheng2016} R. Cheng, D. Xiao, and A. Brataas, Phys. Rev. Lett. {\bf 116}, 207603 (2016).
\bibitem{Johansen} \O. Johansen and J. Linder, Sci. Rep. {\bf 6}, 33845 (2016).
\bibitem{Khymyn} R. Khymyn, I. Lisenkov, V. Tiberkevich, B. A. Ivanov, and A. Slavin, Sci. Rep. {\bf 7}, 43705 (2017).
\bibitem{DKLee} D.-K. Lee, B.-G. Park, and K.-J. Lee, Phys. Rev. Applied {\bf 11}, 054048 (2019).

\bibitem{Gomonay} O. Gomonay, T. Jungwirth, and J. Sinova, Phys. Rev. Lett. {\bf 117}, 017202 (2016).
\bibitem{Ohshba} T. Shiino, S.-H. Oh, P. M. Haney, S.-W. Lee, G. Go, B.-G. Park, and K.-J. Lee, Phys. Rev. Lett. {\bf 117}, 087203 (2016).

\bibitem{SKKim} S. K. Kim, K.-J. Lee, and Y. Tserkovnyak, Phys. Rev. B {\bf 95}, 140404(R) (2017). 
\bibitem{Kabjin} K.-J. Kim, S. K. Kim, T. Tono, S.-H. Oh, T. Okuno, W. S. Ham, Y. Hirata, S. Kim, G. Go, Y. Tserkovnyak, A. Tsukamoto, T. Moriyama, K.-J. Lee, and T. Ono, Nat. Mater. {\bf 16}, 1187-1192 (2017). 

\bibitem{Caretta} L. Caretta, M. Mann, F., B{\"u}ttner, K. Ueda, B. Pfau, C. M. G{\"u}nther, P. Hessing, A. Churikoa, C. Klose, M. Schneider, D. Engel, C. Marcus, D. Bono, K. Bagschik, S. Eisebitt, and G. S. D. Beach, Nat. Nanotechnol. {\bf 13}, 1154-1160 (2018). 
\bibitem{Siddiqui} S. A. Siddiqui, J. Han, J. T. Finley, C. A. Ross, and L. Liu, Phys. Rev. Lett. {\bf 121}, 057701 (2018). 
\bibitem{SHJOM} S.-H. Oh and K.-J. Lee, J. Magn. {\bf 23}, 196 (2018). 

\bibitem{EarlySTT1} A. S. N\'{u}\~{n}ez, R. A. Duine, P. Haney, and A. H. MacDonald, Phys. Rev. B {\bf 73}, 214426 (2006).
\bibitem{EarlySTT2} Z. Wei, A. Sharma, A. S. N\'{u}\~{n}ez, P. M. Haney, R. A. Duine, J. Bass, A. H. MacDonald, and M. Tsoi, Phys. Rev. Lett. {\bf 98}, 116603 (2007).
\bibitem{EarlySTT3} S. Urazhdin and N. Anthony, Phys. Rev. Lett. {\bf 99}, 046602 (2007).
\bibitem{EarlySTT4} P. M. Haney and A. H. MacDonald, Phys. Rev. Lett. {\bf 100}, 196801 (2008).
%
\bibitem{Saidaoui} H. B. M. Saidaoui, A. Manchon, and X. Waintal, Phys. Rev. B {\bf 89}, 174430 (2014).
%
\bibitem{Cheng2014} R. Cheng, J. Xiao, Q. Niu, and A. Brataas, Phys. Rev. Lett. {\bf 113}, 057601 (2014).
%
\bibitem{Xu2008} Y. Xu, S. Wang, and K. Xia, Phys. Rev. Lett. {\bf 100}, 226602 (2008).
%
\bibitem{Swaving2011} A. C. Swaving and R. A. Duine, Phys. Rev. B {\bf 83}, 054428 (2011).
%
\bibitem{Hals2011} K. M. D. Hals, Y. Tserkovnyak, A. Brataas, Phys. Rev. Lett. {\bf 106}, 107206 (2011).
%
\bibitem{Tveten2013} E. G. Tveten, A. Qaiumzadeh, O. A. Tretiakov, and A. Brataas, Phys. Rev. Lett. {\bf 110}, 127208 (2013).
%
\bibitem{Zhang2015arXiv} X. Zhang, Y. Zhou, and M. Ezawa, Sci. Rep. {\bf 6}, 24795 (2016).
%
\bibitem{Barker2015arXiv} J. Barker and O. A. Tretiakov, Phys. Rev. Lett. {\bf 116}, 147203 (2016).
%
\bibitem{Yuta} Y. Yamane, J. Ieda, and J. Sinova, Phys. Rev. B {\bf 94}, 054409 (2016).
%
\bibitem{Yaro} Y. Tserkovnyak, and H. Ochoa, Phys. Rev. B {\bf 96}, 100402(R) (2017).
\bibitem{Tatara} G. Tatara and H. Kohno, Phys. Rev. Lett. {\bf 92}, 086601 (2004).
%
\bibitem{Zhang} S. Zhang and Z. Li, Phys. Rev. Lett. {\bf 93}, 127204 (2004).
%
\bibitem{Thiaville} A. Thiaville, Y. Nakatani, J. Miltat, and Y. Suzuki, Europhys. Lett. {\bf 69}, 990 (2005).
%
\bibitem{Yaro2} Y. Tserkovnyak, H. J. Skadsem, A. Brataas, and G. E. W. Bauer, Phys. Rev. B {\bf 74}, 144405 (2006).
%
\bibitem{Stiles} J. Xiao, A. Zangwill, and M. D. Stiles, Phys. Rev. B {\bf 73}, 054428 (2006).
%
\bibitem{Ohe} J.-i. Ohe and B. Kramer, Phys. Rev. Lett. {\bf 96}, 027204 (2006).
%
\bibitem{Tatara2} G. Tatara, H. Kohno, J. Shibata, Y. Lemaho, and K.-J. Lee, J. Phys. Soc. Japan {\bf 76}, 054707 (2007).
%
\bibitem{Garate} I. Garate, K. Gilmore, M. D. Stiles, and A. H. MacDonald, Phys. Rev. B {\bf 79}, 104416 (2009).

\bibitem{Gilmore1} K. Gilmore, Y. U. Idzerda, and M. D. Stiles, Phys. Rev. Lett. {\bf 99}, 027204 (2007).

\bibitem{Gilmore2} K. Gilmore, I. Garate, A. H. MacDonald, and M. D. Stiles, Phys. Rev. B {\bf 84}, 224412 (2011).
%
\bibitem{Manchon2} A. Manchon and K.-J. Lee, Appl. Phys. Lett. {\bf 99}, 022504 (2011).
%
\bibitem{Manchon} C. A. Akosa, W.-S. Kim, A. Bisig, M. Kl{\"a}ui, K.-J. Lee, and A. Manchon, Phys. Rev. B {\bf 91}, 094411 (2015).
%
\bibitem{Kim2015} K.-W. Kim, K.-J. Lee, H.-W. Lee, and M. D. Stiles, Phys. Rev. B {\bf 92}, 224426 (2015).
%
\bibitem{Hayashi} M. Hayashi, L. Thomas, Ya.B. Bazaliy, C. Rettner, R. Moriya, X. Jiang, and S.S.P. Parkin, Phys. Rev. Lett. {\bf 96}, 197207 (2006).
%
\bibitem{Moriya} R. Moriya, L. Thomas, M. Hayashi, Y.B. Bazaliy, C. Rettner, and S.S.P. Parkin, Nat. Phys. {\bf 4}, 368 (2008).
%
\bibitem{Heyne} L. Heyne, M. Kl{\"a}ui, D. Backes, T.A. Moore, S. Krzyk, U. R{\"u}diger, L.J. Heyderman, A. Fraile Rodr{\`i}guez, F. Nolting, T.O. Mentes, M.{\`A}. Ni{\~n}o, A. Locatelli, K. Kirsch, and R. Mattheis, Phys. Rev. Lett. {\bf 100}, 066603 (2008).
%
\bibitem{Boulle} O. Boulle, J. Kimling, P. Warnicke, M. Kl{\"a}ui, U. R{\"u}diger, G. Malinowski, H.J.M. Swagten, B. Koopmans, C. Ulysse, G. Faini, Phys. Rev. Lett. {\bf 101}, 216601 (2008).
%
\bibitem{Eltschka} M. Eltschka, M. W{\"o}tzel, J. Rhensius, S. Krzyk, U. Nowak, M. Kl{\"a}ui, T. Kasama, R.E. Dunin-Borkowski, L.J. Heyderman, H.J. van Driel, R.A. Duine, Phys. Rev. Lett. {\bf 105}, 056601 (2010).
%
\bibitem{Burrowes} C. Burrowes, A.P. Mihai, D. Ravelosona, J.-V. Kim, C. Chappert, L. Vila, A. Marty, Y. Samson, F. Garcia-Sanchez, L.D. Buda-Prejbeanu, I. Tudosa, E.E. Fullerton, J.-P. Attan{\'e}, Nat. Phys. {\bf 6}, 17 (2010).
%
\bibitem{Sekiguchi} K. Sekiguchi, K. Yamada, S.-M. Seo, K.-J. Lee, D. Chiba, K. Kobayashi, and T. Ono, Phys. Rev. Lett. {\bf 108}, 017203 (2012).
%
\bibitem{Chauleau} J.-Y. Chauleau, H. G. Bauer, H. S. K{\"o}rner, J. Stigloher, M. H{\"a}rtinger, G. Woltersdorf, and C. H. Back, Phys. Rev. B {\bf 89}, 020403(R) (2014).
%
\bibitem{Bisig} A. Bisig, C. A. Akosa, J.-H. Moon, J. Rhensius, C. Moutafis, A. von Bieren, J. Heidler, G. Kiliani, M. Kammerer, M. Curcic,
M. Weigand, T. Tyliszczak, B. Van Waeyenberge, H. Stoll, G. Sch{\"u}tz, K.-J. Lee, A. Manchon, and M. Kl{\"a}ui, Phys. Rev. Lett. {\bf 117}, 277203 (2016).

\bibitem{Okuno} T. Okuno, D.-H. Kim, S.-H. Oh, S. K. Kim, Y. Hirata, T. Nishimura, W. S. Ham, Y. Futakawa, H. Yoshikawa, A. Tsukamoto \textit{et al.},  Nat. Electron. {\bf 2}, 239 (2019).

\bibitem{Balaz} P. Bal{\' a}{\v z}, V. K. Dugaev, and J. Barna{\' s}, Phys. Rev. B. {\bf 85}, 024416 (2012).
\bibitem{Lake} R. Lake, G. Klimeck, R. C. Bowen, and D. Jovanovic, J. Appl. Phys.  {\bf 81}, 7845 (1996).
\bibitem{Datta1} S. Datta, ''Electronic Transport in Mesoscopic Systems'' (Cambridge University Press, Cambridge) 1997.
\bibitem{Ohimp} J. H. Oh, M. Shin, and S.-H. Lee, J. Appl. Phys. {\bf 113}, 233706 (2013).
\bibitem{Kamenev} A. Kamenev and A. Andreev, Phys. Rev. B. {\bf 60}, 2218 (1999).

\bibitem{Landau} L. D. Landau and E. M. Lifshitz, Electrodynamics of Continuous Media, Course of Theoretical Physics Vol. 8 (Pergamon, Oxford, 1960).

\bibitem{Oh} S.-H. Oh, S. K. Kim, J. Xiao, and K.-J. Lee, Phys. Rev. B {\bf 100}, 174403 (2019).

\bibitem{Yamaguchi} A. Yamaguchi, T. Ono, S. Nasu, K. Miyake, K. Mibu, and T. Shinjo, Phys. Rev. Lett.  {\bf 92}, 077205 (2004).

\bibitem{Tveten2016} E. G. Tveten, T. M{\" u}ller, J. Linder, and A. Brataas, Phys. Rev. B {\bf 93}, 104408 (2016).

\bibitem{Tretiakov} O. A. Tretiakov, D. Clarke, G.-W. Chern, Ya. B. Bazaliy, and O. Tchernyshyov, Phys. Rev. Lett. {\bf 100}, 127204 (2008).

\bibitem{Papanicolaou} N. Papanicolaou, Phys. Rev. B {\bf 51}, 15062 (1995).

\bibitem{S.-H.Oh} S.-H. Oh, S. K. Kim, J. Xiao, and K.-J. Lee, Phys. Rev. B. {\bf 100}, 174403 (2019).

\bibitem{Andreev} A. F. Andreev and V. I. Marchenko, Sov. Phys. Usp. {\bf 23}, 21 (1980).

\bibitem{Chiolero} A. Chiolero and D. Loss, Phys. Rev. B {\bf 56}, 738 (1997).

\bibitem{Ivanov} B. A. Ivanov and A. L. Sukstanskii, Solid State Commun. {\bf 50}, 523 (1984).

\bibitem{Aharonov} Y. Aharonov and A. Stern, Phys. Rev. Lett. {\bf 69}, 3593 (1992).

\bibitem{Stiles3} M. D. Stiles and A. Zangwill, J. Appl. Phys. {\bf 91}, 6812 (2002).
\bibitem{Waintal} X. Waintal, E. B. Myers, P. W. Brouwer, and D. C. Ralph, Phys. Rev. B {\bf 62}, 12317 (2000).
\bibitem{Stiles2} M. D. Stiles and A. Zangwill, Phys. Rev. B {\bf 66}, 014407 (2002).

\bibitem{Merodio} P. Merodio, A. Kalitsov, H. Bea, V. Baltz, and M. Chshiev, Appl. Phys. Lett. {\bf 105}, 122403 (2014).

\bibitem{Yang}  J. Yu, D. Bang, R. Mishra, R. Ramaswamy, J. H. Oh, H.-J. Park, Y. Jeong, P. V. Thach, D.-K. Lee, G. Go \textit{et al.}, Nat. Mater. {\bf 18}, 29 (2019)

\bibitem{KJLee} K.-J. Lee, M. D. Stiles, H.-W. Lee, J.-H. Moon, K.-W. Kim, and S.-W. Lee, Phys. Rep. {\bf 531}, 89 (2013).



\bibitem{Boulle2011} O. Boulle, G. Malinowski, M. Kl\''{a}ui, Mater. Sci. Eng. R-Rep. {\bf 72}, 159 (2011).

\bibitem{Kohno2006} H. Kohno, G. Tatara, and J. Shibata, J. Phys. Soc. Jpn. {\bf 75}, 133706 (2006).

\bibitem{Duine2007} R. A. Duine, A. S. N\'{u}\~{n}ez, J. Sinova, and A. H. MacDonald, Phys. Rev. B {\bf 75}, 214420 (2007).

\bibitem{Yaro3} Y. Tserkovnyak, H. J. Skadsem, A. Brataas, and G. E. W. Bauer, Phys. Rev. B {\bf 74}, 144405 (2006).



\end{thebibliography}


\end{document}